\documentclass{elsart}

\usepackage[dvips]{graphicx}

\begin{document}

\newlength{\caheight}
\setlength{\caheight}{12pt}
\multiply\caheight by 7
\newlength{\secondpar}
\setlength{\secondpar}{\hsize}
\divide\secondpar by 3
\newlength{\firstpar}
\setlength{\firstpar}{\secondpar}
\multiply\firstpar by 2

\hfill
\parbox[0pt][\caheight][t]{\secondpar}{
  \rightline
  {\tt \shortstack[l]{
    CERN-TH/99-116 \\
    FNT/T-99/06
  }}
}

\begin{frontmatter}

\title{Light-Pair Corrections to Small-Angle Bhabha \break Scattering
       in a Realistic Set-up at {\tt LEP}}

\author[unipv,infn]{G.~Montagna}
\author[cern,ferrara]{M.~Moretti\thanksref{curie}}
\author[infn,unipv]{O.~Nicrosini}
\author[unipv,infn]{A.~Pallavicini}
\author[infn,unipv]{F.~Piccinini}

\thanks[curie]{Supported by a Marie Curie fellowship ({\sc tmr-erbfmbict} 971934)}

\address[unipv]{Dipartimento di Fisica Nucleare e Teorica\\ Universit\`a di Pavia,
                via A. Bassi 6, Pavia, Italy}
\address[infn]{INFN - Sezione di Pavia, via A. Bassi 6, Pavia, Italy}
\address[cern]{Theory Division, CERN,  CH-1211 Geneva 23, 
               Switzerland}
\address[ferrara]{Dipartimento di Fisica - Universit\`a di Ferrara\\
                  and INFN - Sezione di Ferrara, Ferrara, Italy}

\begin{abstract}
  Light-pair corrections to small-angle Bhabha scattering have been computed
  in a realistic set-up for luminosity measurements at {\tt LEP}.
  The effect of acollinearity and acoplanarity rejection criteria has
  been carefully analysed for typical calorimetric event selections.
  The magnitude of the 
correction, depending on the details of the considered set-up, 
is  comparable with  the present experimental error.
\end{abstract}

\begin{keyword}
                electron$-$positron collision, small-angle Bhabha scattering,
                theoretical error, light pairs, Monte Carlo\\
                {\sc pacs}: 02.70.Lq,12.15.Lk,13.40.Ks,13.85.Hd   
\end{keyword}

\vfill

\end{frontmatter}

\newpage
\begin{table}
\caption{Current status of the theoretical error for the {\tt SABH} scattering,
\newline
         as reported in refs.~\cite{pairs,ward,lmw}.} \label{sabherr}
  \medskip
  \begin{center}
  \begin{tabular}{|l||c|c|} \hline
    & \multicolumn{2}{c|}{Uncertainty \cite{pairs,ward,lmw}}\\ \cline{2-3}
    Type of correction/error & {\tt LEP1} ($\%$) & {\tt LEP2} ($\%$) \\ \hline \hline
    Missing photonic $O(\alpha^2L)$              & $0.027 $ & $0.040$\\
    Missing photonic $O(\alpha^3L^3)$            & $0.015 $ & $0.030$\\
    Vacuum polarization                          & $0.040 $ & $0.100$\\
    Light pairs                                  & $0.010 $ & $0.015$\\
    $Z$-exchange                                 & $0.015 $ & $0.000$\\ \hline
    Total error                                  & $0.054 $ & $0.113$\\ \hline
  \end{tabular}
  \end{center}
\end{table}

In the last years, many efforts were made to reduce the
sources of theoretical error in the prediction of the small-angle
Bhabha  (hereafter {\tt SABH}) scattering cross section,
in order to match the increased experimental accuracy. The main
results were achieved in the sector of the $O(\alpha^2L)$ photonic 
corrections,
by lowering the associated uncertainty to the $0.03\%$ level \cite{kr,pv}.
Moreover, the uncertainty associated to the light-pair contribution was reduced
\cite{russi,jadach,pairs} to the $0.01\%$ level.
The ultimate result of these works was to lower the total theoretical
error to the $0.05\%$ level for {\tt LEP}1 energies, as it can be read from
table~\ref{sabherr} (see also refs.~\cite{ward,lmw}).
On the other hand, at present, the experimental error associated to
luminosity measurements is below the $0.05\%$ level. Since the size 
of light-pair contributions is of the order of some 
$0.01\%$~\cite{russi,jadach,pairs} and 
will depend, in general, on the event selection (hereafter  ES),
it is important to include the best available estimate for light-pair
corrections. In particular, the presence of tight cuts, which select
events with soft-pair emission, such as acollinearity and acoplanarity
cuts, can significantly alter the light-pair correction.

At present, the theoretical error to {\tt SABH} scattering, due to pair
production, can be evaluated by approximate means, such as the 
Monte Carlo (hereafter MC)
results based on $t$-channel approximation \cite{jadach} or the analytical
calculations in the quasi-collinear approximation \cite{russi}, or
by the MC calculation of ref.~\cite{pairs}, which
includes the exact QED four-fermion matrix element,
two-loop virtual corrections according to ref.~\cite{burgers},
initial-state radiation (hereafter ISR) in the collinear
approximation, and realistic  ES's.
In this note the light-pair contribution is studied in the presence of  ES's
as realistic as possible by using the approach of ref.~\cite{pairs},
to which the reader is addressed for any technical detail.
Particular attention is payed to the {\tt OPAL} ES \cite{opal},
as a significant case study.

Before entering the details of the {\tt OPAL}  ES, it is worthwhile to 
 consider
a typical calorimetric ES, such as one of the {\tt CALO2}  ES's
adopted in ref.~\cite{pairs}.
Apart from other technical features, it is characterized by an energy
cut defined in terms of the kinematical variable $z$:

\begin{figure}
  \begin{center}
    \raisebox{-16ex}{\includegraphics[bb=55 385 280 610,scale=.6]
    {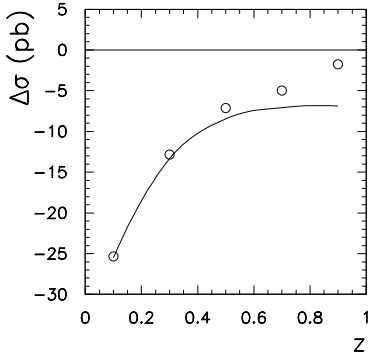}}
  \end{center}
\caption{The MC integration of the exact matrix element (markers) 
         \cite{pairs} and the $t$-channel approximation recalculated according to
         \cite{jadach} (solid line) as a function of the energy cut
         $z\equiv 1-{E_{\rm 1}E_{\rm 2}/ E^2_{\rm beam}}$.
         Entry values are in pb and sum up real and virtual corrections; they
         are computed for the {\tt CALO2} set-up with asymmetric acceptance
         $3.49^\circ\leq\theta_{\rm N}\leq 6.11^\circ$,
         $2.97^\circ\leq\theta_{\rm W}\leq 6.73^\circ$ at $\sqrt{s}=92.0{\rm~GeV}$.
         They include only the electron contribution without ISR.
         In this set-up, the tree-level Bhabha  cross section is $\sigma_0=21939(1) \;\mathrm{pb}$.
        } \label{vary}
\end{figure}

\begin{equation}
  z \equiv 1-{E_{\rm 1}E_{\rm 2}\over E^2_{\rm beam}} \leq z_{\rm max} ,
\end{equation}

\noindent where $E_{\rm 1,2}$ are the energies of the two  clusters of 
particles
hitting the forward and backward calorimeters. The definition of the ES,
involving angular and energy cuts only, is reported in the caption of fig.~\ref{vary}.
Notice that small values of $z$ inhibit hard-pair emission, while large values
do not. In fig.~\ref{vary} the light-pair contribution to {\tt SABH} scattering
is shown 
as a function of $z$. As expected, the magnitude of the correction
undergoes a significant variation by changing the $z$ value. In particular
the correction grows in absolute value if the available phase-space region
favours soft-pair radiation. This is the same behaviour as 
observed if one studies
photon emission, instead of pair emission.
It is worth noticing that the enhancement of the pair correction is valid within
the $t$-channel approximation too.
It is also important to stress that superimposing an acollinearity
and acoplanarity cut means inhibiting hard radiation, so that such a cut can
effectively mimic a cut on $z$, constraining it in the soft region.
As an example, these cuts were superimposed on {\tt CALO2},
by considering only the electron-pair contribution at the tree level.
The results are shown in table~\ref{caloplus}
for asymmetric angular acceptances ($3.49^\circ\leq\theta_{\rm N}\leq 6.11^\circ$ and
$2.97^\circ\leq\theta_{\rm W}\leq 6.73^\circ$) at $\sqrt{s}=92.0{\rm~GeV}$
with acollinearity and acoplanarity cuts ($\theta_{\rm ac}=0.58^\circ$
and $\phi_{\rm ap}=11.46^\circ$).
This exercise shows that the presence of acollinearity and acoplanarity cuts
increases, in absolute value, the light-pair correction significantly, i.e.
it has the same effect of lowering the $z$ cut.\footnote{With the given values
of the acollinearity and acoplanarity cuts, the largely dominant effect is due
to the acollinearity cut.} This link can be easily understood
since acollinearity and acoplanarity cuts select events with
 soft-pair radiation.
It is worth noticing that the higher the value of $z$, the higher the relative
enhancement of the correction with respect to the correction itself, as expected.

\begin{table}
\caption{Comparison between {\tt CALO2} asymmetric  ES
         with and without a further cut in acollinearity and acoplanarity
         ($\theta_{\rm ac}=0.58^\circ$, $\phi_{\rm ap}=11.46^\circ$) at
         $\sqrt{s}=92.0{\rm~GeV}$. {\tt CALO2} set-up acceptances
         are $3.49^\circ\leq\theta_{\rm N}\leq 6.11^\circ$ and
         $2.97^\circ\leq\theta_{\rm W}\leq 6.73^\circ$.
         First-column entries are in pb and refer to the MC
         integration of the exact matrix element without ISR
         at $z=0.3,~0.5,~0.7$, where $z\equiv 1-{E_{\rm 1}E_{\rm 2}/ E^2_{\rm beam}}$,
         and sum up real and virtual part. Second-column entries
         refer to the relative correction with respect to the Bhabha
         tree-level cross section $\sigma_0=21939(1) \;\mathrm{pb}$.
        } \label{caloplus}
  \medskip
  \begin{center}
    \begin{tabular}{|l|c||c|c|} \hline
      Set-up modality           & $z$   & Abs.~corr.~(pb)    & Rel.~corr.~($10^{-4}$) \\ \hline \hline
      {\tt CALO2}              & $0.3$ & $-12.85 \pm 0.05$ & $ -5.86 \pm 0.02$ \\
      {\tt CALO2} with ac./ap. &       & $-18.68 \pm 0.06$ & $ -8.51 \pm 0.03$ \\ \hline
      {\tt CALO2}              & $0.5$ & $ -7.14 \pm 0.05$ & $ -3.25 \pm 0.02$ \\
      {\tt CALO2} with ac./ap. &       & $-16.36 \pm 0.19$ & $ -7.46 \pm 0.08$ \\ \hline
      {\tt CALO2}              & $0.7$ & $ -4.98 \pm 0.12$ & $ -2.27 \pm 0.06$ \\
      {\tt CALO2} with ac./ap. &       & $-15.31 \pm 0.10$ & $ -6.98 \pm 0.05$ \\ \hline
    \end{tabular}
  \end{center}
\end{table}

Let us now consider the more realistic {\tt OPAL} case.
The {\tt OPAL} luminosity is measured with an experimental precision of
$0.034\%$ \cite{opal}, and similar performances are attained by the other
collaborations. On the other hand, the light-pair correction is of the order
of some $0.01\%$ and, moreover, it could be critically enhanced by
tight acollinearity and acoplanarity cuts, as just shown in the {\tt CALO2}
case (see table \ref{caloplus}). It is thus crucial, 
for the luminosity measurements,
to include a careful estimate of the pair corrections.
The {\tt OPAL} collaboration defines a reference theoretical cross section in terms of
simple cuts at four-vector level on the generated particles \cite{opal}.
This rejection set-up, reviewed in table \ref{m4sel}, is named {\tt M4SEL};
it comes in three flavours {\tt SWITL}, {\tt SWITR} and {\tt SWITA},
corresponding to whether the narrow cut in polar angle is applied to the
left or the right hand calorimeter, or, in the case of 
{\tt SWITA}, to the
average of the polar angles measured on the right and left \cite{opal}.

\begin{table}
\caption{{\tt OPAL} rejection set-up defining the reference theoretical
         cross section for the luminosity acceptance calculation} \label{m4sel}
  \medskip
  \begin{center}
    \begin{tabular}{|l||c|} \hline
      Wide                 &  $1.56^\circ$ $-$  $3.19^\circ$ \\
      Narrow               &  $1.79^\circ$ $-$ $2.96^\circ$ \\
      Max acollinearity    &  $0.58^\circ$ \\
      Max acoplanarity     & $11.46^\circ$ \\
      Min energy per calo  &  $\sqrt{s}/4$ \\
      Min energy total     & $3\sqrt{s}/4$ \\ \hline
    \end{tabular}
  \end{center}
\end{table}

\begin{table}
\caption{Light-pair correction to {\tt SABH} scattering. Entry values sum up
         real and  virtual corrections, and are computed for the {\tt M4SEL}
         set-up at $91.0{\rm~GeV}$. The first column shows the absolute correction
         in pb, while the second column shows the correction relative to the
         Born cross section $\sigma_0=81344(7) \;\mathrm{pb}$. 
         The errors quoted sum up the physical and the
         technical error as estimated according to ref.~\cite{pairs}.} \label{opallp}
  \medskip
  \begin{center}
    \begin{tabular}{|l||c|c|} \hline
      Set-up modality & Abs.~corr.~(pb)  & Rel.~corr.~($10^{-4}$) \\ \hline \hline
      {\tt SwitL} & $-35.99 \pm 11.55$ & $-4.42 \pm 1.43$ \\
      {\tt SwitR} & $-35.83 \pm 11.33$ & $-4.40 \pm 1.40$ \\
      {\tt SwitA} & $-35.64 \pm 11.36$ & $-4.38 \pm 1.40$ \\ \hline
    \end{tabular}
  \end{center}
\end{table}

A complete calculation, performed with the  MC code of ref.~\cite{pairs},
including ISR via collinear structure functions and the muons contribution,
gives the results shown in table \ref{opallp}, leading to a correction
of $-0.044\%$.
Two comments are in order here. The first is that the pair correction
is of the same order as the experimental error. The second is that the pair
correction computed for an  ES with similar angular and energy cuts,
but without an acollinearity and acoplanarity cut, is at the level of
$-0.025$ -- $0.030\%$ \cite{pairs}, i.e. smaller than the present prediction.

In this short discussion the relevance of taking into account
the light-pair correction to {\tt SABH} scattering for luminosity
measurements at $e^+e^-$ colliders is pointed out. This need is due to
the high experimental accuracy now achieved by the {\tt LEP}
collaborations, better than the $0.05\%$ level. In particular the
effects of acollinearity and acoplanarity cuts are analysed, and
general arguments are given to understand why the presence of such rejection
criteria increases the size of light-pair corrections.
Moreover a realistic  ES, the {\tt M4SEL} adopted by the {\tt OPAL}
collaboration, has been implemented in a MC code to size the light-pair
contribution to {\tt SABH} scattering, leading to a correction for light
pairs at the $0.04\%$ level.

\vspace*{2ex}
\noindent{\bf Acknowledgements}

\vspace*{2ex}
\noindent The authors wish to thank R.~Kellogg and D.~Strom, of the {\tt OPAL} collaboration
for several stimulating and fruitful discussions and for having provided the specifications
of the {\tt OPAL} reference ES of table \ref{m4sel}.

\renewcommand{\refname}{{\normalsize \bf References}}

\end{document}